\documentstyle[amssymb,preprint,prl,aps,12pt]{revtex}
\begin{document}
\title{{\bf Dynamics of human walking}}
\author{Valery B. Kokshenev}
\address{{\it Departamento de F\'{i}sica, Universidade Federal de Minas Gerais,}\\
ICEx, Caixa Postal 702, CEP 30123-970, Belo Horizonte, MG, Brazil}
\date{\today }
\maketitle

\begin{abstract}
\leftskip 54.8pt \rightskip 54.8pt

The problem of biped locomotion at steady speeds is discussed through a
Lagrangian formulation developed for velocity-dependent, body driving
forces. Human walking on a level surface is analyzed in terms of the data on
the resultant ground-reaction force and the external work. It is shown that
the trajectory of the center of mass is due to a superposition of its
rectilinear motion with a given speed and a backward rotation along a
shortened hypocycloid. A stiff-to-compliant crossover between walking gaits
is described and the maximum speed for human walking, given by an
instability of the trajectory, is predicted.


\end{abstract}


In biology, many fundamental discoveries come from studies of animal
movement. The integrative approach to the locomotion of animals focuses on
the interaction between the muscular, tendon and skeletal subsystems and the
environment.\cite{DFF00}

During locomotion, the body as a whole performs several functions. Chemical
energy released by muscles and mechanical elastic energy stored in passive
muscles and tendons\cite{RMW97} are transformed into external and internal
work,\cite{WCH95} and are partially lost as a heat. The muscular system
provokes the ground-reaction forces applied to the animal body. The
resultant force, including gravity and air resistance, accelerates and
decelerates the body's center of mass (COM). This {\em body driving force}
is therefore involved in level walking and running, even when the average
velocity remains constant. Over a complete step cycle, the driving force
performs a certain {\em external work} to maintain a given speed. Studies of
the mechanical efficiency\cite{WCH95} of animal locomotion at different {\em %
steady speeds}, determined through the external work and oxygen consumption,
provide evidence that walking is more energetically economical than running.
This finding corroborates the old idea that walking in humans, primates, and
ground-dwelling birds can be understood as swings of an ideal pendulum.
Indeed, the body vaults up and over each leg in an arc in each step,
similarly to an inverted plane pendulum, and kinetic energy is transformed
into gravitational energy when the body falls forward and downward. Such a
stiff-legged mechanics of walking modeled by the compass-arc inverted
pendulum is widely employed,\cite{DFF00,WCH95,CM02} but by no means
exhaustive.\cite{DFF00,Sch03} Unlike the swing pendulum dynamics driven by
constant gravitational force, the dynamics of animal biped walking (and
quadruped trotting) is accompanied by body undulations, pulses, and
peristaltic waves\cite{DFF00} generated by the ground-reaction force. In
this Letter, we employ the fundamental principles of classical mechanics to
approach to the problem of level locomotion. At very low speeds, animal
movement is treated through the linear vibrations of the body near its
quasistatic equilibrium given by quiet standing.\cite{LCP98} Nonlinear body
motion effects, controlled by velocity-dependent forces, are taken into
account at higher speeds.

All the three components (forward, vertical and lateral) of the resultant
force applied to the ground are measured with good accuracy by means of the
force-platform techniques.\cite{Cav75} The lateral body displacements are
relatively small and, thus, the COM motion can be fairly described by the
instantaneous polar vector ${\bf R}(t)$ defined in the ground coordinate
system (see {\bf Fig. 1}). For walking at a steady speed $V$, it is
convenient to exclude the translational degree of freedom by introducing $%
{\bf r}(t)$, with $x(t)=Vt$ and $y(t)=H$, where $H$ is the height of the
COM. In that way, the libration motion is given by $\Delta {\bf r}(t)={\bf R}%
(t)-{\bf r}(t)$. The corresponding driving force $\Delta {\bf F}(t)$ follows
from the force-platform records: the ground-reaction force ${\bf F}(t)$ is
observed as the oscillating force near the body weight, thus, ${\bf F}(t)=-m%
{\bf g}+\Delta {\bf F}(t)$. Taking into account that the muscle-tendon
contractions are cyclic, the driving force must satisfy the {\em %
steady-motion constraint}: $<\Delta {\bf F}(t)>_{c}\equiv
T_{c}^{-1}\int_{0}^{T_{c}}$ $\Delta {\bf F}(t)dt=0$, where $T_{c}(V)$ is the
one-step ground-contact period (shown in Fig. 1).

Assuming the displacements $\Delta {\bf r}$ to be small, we introduce the
librational part of the potential energy $\Delta U[\Delta {\bf r}(t)]$ in
the harmonic approximation: $\Delta U_{0}=k_{0}(\Delta x^{2}+\Delta y^{2})/2$
through the {\em body stiffness coefficient} $k_{0}(V)$. Combining this with
the kinetic energy $\Delta K_{0}=m(\Delta \stackrel{\bullet }{x}^{2}+\Delta 
\stackrel{\bullet }{y}^{2})/2$ and employing Lagrangian formalism, one
deduces the Newton equations $m\Delta \stackrel{\bullet \bullet }{\bf r}%
+k_{0}\Delta {\bf r}=0$. The free COM motion is therefore a superposition of
the two linear oscillations: 
\begin{equation}
\Delta x_{0}(t)=\Delta l_{0}\cos (\omega _{0}t-\frac{\pi }{2})\text{; }%
\Delta y_{0}(t)=\Delta h_{0}\sin (\omega _{0}t-\frac{\pi }{2})\text{.}
\label{delta-r0}
\end{equation}
These are solutions $\Delta {\bf r}_{0}{\bf (}t{\bf )}$ given by the
harmonic amplitudes $\Delta l_{0}(V)$ and $\Delta h_{0}(V)$ and the one-step 
{\em angular frequency} $\omega _{0}(V)$, with $\omega _{0}=2\pi /T_{c}=%
\sqrt{k_{0}/m}$. The backward elliptical COM rotation in Eq.(\ref{delta-r0})
is due to the harmonic part of the driving force $\Delta {\bf F}%
_{0}(t)=-k_{0}\Delta {\bf r}_{0}$, derived from the experiment and treated
as an inertial force. Its components:

\begin{equation}
\Delta F_{0x}(t)=-m\omega _{0}^{2}\Delta l_{0}\sin (\omega _{0}t)\text{; }%
\Delta F_{0y}(t)=m\omega _{0}^{2}\Delta h_{0}\cos (\omega _{0}t),
\label{delta-F0}
\end{equation}
are shown by solid lines in Fig. 1. With increasing speed, anharmonic
displacements become important and therefore $\Delta U=\Delta U_{0}+\Delta
U_{1}$. Without loss of generality, the anharmonic part of the mechanical
potential energy $\Delta U_{1}$ is parametrized in terms of the anharmonic
force amplitudes $\Delta l_{1}$ and $\Delta h_{1}$, namely

\begin{equation}
\Delta U_{1}(\Delta {\bf r})=\frac{k_{0}}{\Delta h_{0}}\left[ -\frac{\Delta
l_{1}}{\Delta l_{0}}\Delta x^{2}\Delta y+\frac{\Delta h_{1}}{3\Delta h_{0}}%
\Delta y^{3}+O(\Delta x\Delta y^{2})+O(\Delta x^{3})\right] .  \label{U1}
\end{equation}
Within the perturbation scheme, the nonlinear forces are defined by the
derivatives $\Delta {\bf F}_{1}=-d\Delta U_{1}/d\Delta {\bf r}$ taken at $%
\Delta {\bf r}=\Delta {\bf r}_{0}$ given in Eq.(\ref{delta-r0}). This
results in 
\begin{eqnarray}
F_{x}(t) &=&-m\omega _{0}^{2}[\Delta l_{0}\sin (\omega _{0}t)+\Delta
l_{1}\sin (2\omega _{0}t)]\text{;}  \nonumber \\
F_{y}(t) &=&mg+m\omega _{0}^{2}[\Delta h_{0}\cos (\omega _{0}t)-\Delta
h_{1}\cos (2\omega _{0}t)].  \label{Fx-Fy}
\end{eqnarray}
The third and the fourth terms in Eq.(\ref{U1}) correspond to the terms of $%
O[\cos (2\omega _{0}t)]$ and of $O[\sin (2\omega _{0}t)]$, which formally
should appear in $F_{x}$ and $F_{y}$, respectively. Both the terms are
omitted in Eqs.(\ref{Fx-Fy}) because they are not observed in the available
data on human walking.\cite{Cav75,CHT77} Also, the steady-motion constraint
provides the force-amplitude relation $\Delta l_{0}\Delta l_{1}=\Delta
h_{0}\Delta h_{1}$ that is nevertheless violated, even in the case of the
small $V$. As a matter of fact, the theory behind this relation presumes
that $\Delta {\bf F}_{1}$ is a conservative force, which disagrees with the
experimental data.

Let us introduce a generalized velocity-dependent Lagrangian\cite{GOLD80} $%
\Delta L(\Delta {\bf r,}\Delta \stackrel{\bullet }{\bf r})=\Delta
K_{0}-\Delta U_{eff}$, where $\Delta U_{eff}=\Delta U_{0}+\Delta
U_{1}-\Delta K_{1}$ and $\Delta K_{1}$ is the anharmonic kinetic energy.\cite
{LIL} Within the scope of this analysis, Eqs. (\ref{Fx-Fy}) are not altered
and the steady-motion constraint is satisfied by new kinetic terms. With the
help of the {\em frictional coefficient} $\gamma (V)$, we also introduce the
resistance force $\Delta {\bf F}_{res}(t)=-\gamma $ $\Delta \stackrel{%
\bullet }{\bf r}_{1}$, associated with the anharmonic displacements $\Delta 
{\bf r}_{1}\equiv \Delta {\bf r}-\Delta {\bf r}_{0}$. The latter obey the
equations 
\begin{equation}
m\Delta \stackrel{\bullet \bullet }{\bf r}_{1}+\gamma \Delta \stackrel{%
\bullet }{{\bf r}_{1}}+k_{0}\Delta {\bf r}_{1}=\Delta {\bf F}_{1}(t)\text{,}
\label{Newton eq}
\end{equation}
where $\Delta {\bf F}_{1}(t)$ is given by the last terms in Eqs. (\ref{Fx-Fy}%
). Solutions of inhomogeneous differential equations (\ref{Newton eq})
provide the desired description for the body's COM motion in the ground
coordinate system:

\begin{eqnarray}
X(t) &=&Vt+\Delta x_{0}(t)+\frac{\Delta l_{1}}{3}\frac{\sin (2\omega
_{0}t+\varphi )}{\sqrt{1+\tan (\varphi )^{2}}}\text{, }  \nonumber \\
Y(t) &=&H+\Delta y_{0}(t)+\frac{\Delta h_{1}}{3}\frac{\cos (2\omega
_{0}t+\varphi )}{\sqrt{1+\tan (\varphi )^{2}}}\text{.}  \label{delta-x-y}
\end{eqnarray}
They are found\cite{GOLD80,LIL} for the steady motion regimes in the weak
friction approximation, with 
\begin{equation}
\text{ }\varphi (V)=\arctan \left( \frac{2\omega _{1}}{3\omega _{0}}\right) <%
\frac{\pi }{4}\text{; }\omega _{1}(V)=\frac{\gamma }{m}\text{. }
\label{param}
\end{equation}
As seen from Eqs.(\ref{delta-x-y}), these regimes are established not only
by the frequencies, but also by the amplitudes. The latter statement follows
from the force-time fitting analysis given in Fig. 1 that can be explicit in
the force-amplitude ratio: $\Delta l_{0}^{(\exp )}/\Delta l_{1}^{(\exp
)}=\Delta h_{0}^{(\exp )}/\Delta h_{1}^{(\exp )}$ which equals $2$. As can
be recognized from Eqs. (\ref{delta-r0}), (\ref{delta-x-y}) and the
force-amplitude ratio, a trajectory of the human body's COM, in the moving
inertial coordinate system, is a closed orbit given by a shortened
hypocycloid ($\Delta r_{1}<\Delta r_{0}\ll H$), passing through three
turning points in the backward direction.\cite{exe} For qualitative
analysis, we describe this closed orbit by the characteristic ellipse, which
crosses the same turning points and is introduced by the axes: $\Delta l=$ $%
\Delta l_{0}+\Delta l_{1}/3\sqrt{1+\tan (\varphi )^{2}}$ and $\Delta
h=\Delta h_{0}+\Delta h_{1}/3\sqrt{1+\tan (\varphi )^{2}}$ , in the forward
and vertical directions, respectively. One infers that the COM moves on the
height $H$ with the speed $V$, and simultaneously rotates along the
hypocycloid circumscribed by a shrunken (or a flattened) ellipse of
eccentricity $e_{+}$ (or $e_{-}$), with $e_{\pm }(V)=\sqrt{1-(\Delta
l_{1}/\Delta h_{1})^{\pm 2}}$ .

The step-cycle external work $W_{tot}${\em \ }is performed by the COM to
maintain the forward speed $V$ and the height $H$, thus, $%
W_{tot}=W_{f}+W_{v} $. This work can be estimated through the instant power
averaged over the cycle period: $W_{tot}(V)=2\pi \omega _{0}^{-1}<\stackrel{%
\bullet }{W}_{tot}(t)>_{c}$, where $\stackrel{\bullet }{W}_{tot}(t)={\bf %
F\cdot }\stackrel{\bullet }{\bf R}$. Nevertheless, not all the components of
the ground-reaction force, produced by active and passive muscles,
contribute to the cyclic work. Bearing in mind the conditions i) of the
periodicity of the driving force, ii) of the orthogonality between the
linear and nonlinear displacements, and iii) of the conservative nature of
the harmonic force, one deduces that the only nonzero contribution to $%
W_{tot}$ is due to the anharmonic part of the power $\stackrel{\bullet }{W}%
_{1}(t)=\Delta {\bf F}_{1}\Delta \stackrel{\bullet }{\bf r}_{1}$. This power
follows from Eqs.\ (\ref{delta-x-y}) and provides 
\begin{equation}
W_{tot}(V)=\frac{4\pi }{9}\frac{\gamma \omega _{0}(\Delta l_{1}^{2}+\Delta
h_{1}^{2})}{1+(2\gamma /3m\omega _{0})^{2}}\text{.}  \label{w-tot}
\end{equation}
On the one hand, the total external work corresponds to that part of the
mechanical energy that is lost as heat. It must be therefore restored in the
next step through chemical energy by oxygen consumption. On the other hand,
the external work is realized through the positive and negative
contributions: $W_{tot}=W^{+}-|W^{-}|$. These can be exemplified by the
accelerated and decelerated forward body's displacements, respectively. A
special case is ideal oscillational motion, when the two contributions are
equal and, thus, $W_{tot}=0$. Also, the {\em recovery coefficient}, defined%
\cite{WCH95} for arbitrary cyclic motion as $r=(W^{+}-W_{tot})/W^{+}$,
equals one. To estimate $r(V)$, we specify the positive work by $%
W^{+}(V)=T_{c}<P[\stackrel{\bullet }{W}_{1}(t)]>_{c}$, where the auxiliary
function $P(x)=xH(x)$, with $H(x)$ is the Heavyside step function. This
leads to the recovery coefficient 
\begin{equation}
r(\omega _{0},\omega _{1})=1-\frac{\left( \overline{P[\cos (2\omega
_{0}t)\sin (2\omega _{0}t+\varphi )]}\right) ^{-1}}{2\sqrt{(3\omega
_{0}/2\omega _{1})^{2}+1}}\text{.}  \label{rec}
\end{equation}
If one employs the human-walk data on $\omega _{0}^{(\exp )}(V)$\cite{GB91}
and $r^{(\exp )}(V)$,\cite{WCH95} Eq.(\ref{rec}) can be read as $r(\omega
_{0}^{(\exp )},\omega _{1})=r^{(\exp )}$ and solved for the mass-specific
frictional coefficient (see {\bf Fig. 2}). Additionally, a straightforward
estimation of the positive work performed in the forward ($W_{f}^{+}$) and
vertical ($W_{v}^{+}$) directions, yields the anharmonic amplitudes

\begin{equation}
\Delta l_{1}(V)=\sqrt{\frac{W_{f}^{+}(1-r)}{8\pi m}\frac{9\omega
_{0}^{2}+4\omega _{1}^{2}}{\omega _{0}^{3}\omega _{1}}}\text{; }\Delta
h_{1}(V)=\Delta l_{1}\sqrt{\frac{W_{v}^{+}}{W_{f}^{+}}}\text{.}
\label{amplitudes}
\end{equation}

As follows from a numerical analysis of Eqs.(\ref{amplitudes}) given in {\bf %
Fig. 3}, the instability of the vertical COM librations occurs at the
maximum speed for human walking with $V_{\max }=3.4$ $m/s$, in accord with
the recent experimental data\cite{Sch03} $V_{\max }^{(\exp )}=3.2$ $m/s$.
The eccentricity of the orbit-characteristic ellipses (see {\bf Fig. 4})
specifies this instability by the critical condition $e_{-}(V_{\max })=1$
associated with a dynamical transition from walking to running. There is
also a dynamical crossover at the speed $V_{cr}=1.7$ $m/s$, which separates
slow walking from fast walking. At this speed, $e_{\pm }(V_{cr})=0$ and the
shrunken ellipses ($\Delta l<\Delta h$) transform into the flatter ellipses (%
$\Delta h<\Delta l$). In reality, the crossover in locomotion is attributed
to changes in performance of the human legs in the stiff-legged ($\Delta
l<\Delta h$) and the compliant ($\Delta h<\Delta l$) walking. These two
walking gaits are distinguished through distinct postures (and
reaction-force records) and identified with a modern human walk and a walk
of nonhuman primates, respectively.\cite{Sch03}

In conclusion, we have discussed the problem of animal locomotion on level
ground in view of human data on reaction force and external work for walking
at steady speeds. Application of the standard Lagrangian formalism developed
for nonconservative forces, permits one to introduce a generalized equation
for animal locomotion in the point-body-mass approximation. Such a
description involves only integrative properties of body, given through the
inertial, elastic, and resistance characteristics, and is therefore expected
to be helpful for comparative studies of quadruped trotting of animals where
lateral effects are also negligible. For the case of human walking, the
locomotion for the body's COM\ is a superposition of the rectilinear motion
with its backward rotation along a shortened hypocycloid.

Records of the ground-reaction force elucidate a variety of body's
functions. In animal locomotion, the observed reaction force acts as a
motor-brake force, which additionally supports the body weight and controls
the stability of forward advancement. The efficiency of the employed above
body-support function is restricted by anatomical adaptation of the
long-bone limbs for peak force-ground contacts. This adaptation was proven%
\cite{KSG03} to be universal for all terrestrial mammals, with evolution of
their body mass. As to motor-brake effects, we have deduced that the
velocity-independent forces, which produce harmonic body vibrations, are
responsible for an effective-exchange mechanism of elastic mechanical
energy, attributed to passive muscles and parametrized by the body stiffness
coefficient. The nonlinear anharmonic librations, which are due to the
nonconservative part of the driving force, produce the main part of the
external work. The feedback between linear and nonlinear COM librations is
revealed through the phase, the amplitude and the frequency constraints
imposed on cyclic human walking. Finally, our analysis of the external work
given without recourse to a swing-pendulum or spring-mass modeling provides
insights into the two principal gaits of biped walking. We have seen that
the stiff-to-compliant crossover in human walking arises from the changes of
body-resistance performance and not from the body-elastic adjustment,
prescribed earlier by the pendulum dynamics.

The author is grateful to Ronald Dickman for helpful comments and advice in
preparation of this manuscript. Financial support by CNPq is also
acknowledged.

\begin{center}
{\Large FIGURE\ CAPTURES}
\end{center}

Fig. 1. Analysis of force-platform data on human level walking. Records for
the horizontal ($F_{x}$) and vertical ($F_{y}$) forces are taken from Fig. 1
in Ref.\cite{CHT77} (the case of mass $59$ $kg$ and of speed $3.9$ $km/h$)
and given for the two-step period $2T_{c}$. Solid lines describe the
force-time linear harmonics at distinct phases (denoted by letters) explicit
in Eq.(\ref{delta-F0}) along with the fitting amplitudes $\Delta l_{0}=0.012$
$m$ and $\Delta h_{0}=0.016$ $m$. The lines indicated by hatched areas
correspond to the double-frequency harmonics given in Eqs.(\ref{Fx-Fy}) and
adjusted with $\Delta l_{1}=0.006$ $m$ and $\Delta h_{1}=0.008$ $m$; $H$ ($%
\thickapprox 1m$) stands for the body's COM\ height. Asymmetric deviations
are due to the differences between i) the exertions by left and right feet
and ii) the time intervals for the single-foot ($T_{sc}$) and double-foot ($%
T_{dc}$) ground contacts, with $T_{c}=$ $T_{sc}+T_{dc}$. Rotations of the
driving force and the corresponding velocity are shown by dotted lines in
the coordinates $\Delta F_{y}(t)$ {\it vs} $\Delta F_{x}(t)$ and $\Delta
V_{y}(t)$ {\it vs} $\Delta V_{x}(t)$, respectively.

Fig. 2. Human-walk characteristic frequencies against steady speed. Points
for the cyclic frequency $\omega _{0}(V)$ reproduce cinematographic data
(open circles) taken from Fig. 3 in Ref.\cite{GB91}. They are extended by
the case analyzed in Fig. 1 (open square) and fitted by $\omega
_{0}(V)=4.94+4.02V$ (solid line). Points for the mass-specific frictional
coefficient $\gamma (V)/m$ (shaded circles) are found through Eq.(\ref{rec})
with the help of $\omega _{0}(V)$ and $r^{(\exp )}(V)$ (reproduced in the
insert from Fig. 2 in Ref.\cite{WCH95}) and fitted by $\omega
_{1}(V)=6.37-6.15V+2.38V^{2}$ (given by the solid curve).

Fig. 3. Anharmonic amplitudes for the driving force against speed in human
walking. The horizontal and vertical amplitudes $\Delta l_{1}(V)$ and $%
\Delta h_{1}(V)$, reduced by $\Delta l_{0}(1.1)=0.012$ $m$, are given by
open and shaded circles, respectively. They are estimated through Eqs.(\ref
{amplitudes}) with the help of the parameters obtained in Fig. 2 and of data
on the positive work $W_{f}^{+(\exp )}$ and $W_{v}^{+(\exp )}$ (taken from
Fig. 2 in Ref.\cite{WCH95}). The positive forward and vertical works are
adjusted at $V=1.1$ $m/s$ (shown by open and shaded squares, respectively).
The curves are third-order polynomial fits extrapolated to the maximum speed 
$V_{\max }=3.4$ $m/s$ (shown by the arrow).

Fig. 4. The human center-of-mass body orbit characteristics for slow and
fast level walking. Inserts show ellipses, which circumscribe the closed COM
trajectories in the inertial coordinate system moving with different speeds $%
V$ (indicated by arrows). The elliptic axes are reduced by the amplitude $%
\Delta l_{0}=0.012$ $m$. Experimental data on the elliptic eccentricities
(shown by open circles) for the shrunken and flatter ellipses obtained,
respectively, through $\sqrt{1-W_{f}^{+(\exp )}/W_{v}^{+(\exp )}}$ and $%
\sqrt{1-W_{v}^{+(\exp )}/W_{f}^{+(\exp )}}$, with the help of the
experimental data reported in Ref.\cite{WCH95}. The fitting curve
corresponds to the analysis given in Fig. 3.


\begin{references}
\bibitem{exe}  The hypocycloid with the three turning points can be
illustrated by a trajectory of the driving force, shown by the dotted line
in Fig. 1.
\end{references}
\end{document}